\documentclass{ws-procs975x65}

\begin{document}

\title{\uppercase{Bose-Einstein condensate dark matter model tested by galactic rotation curves}}

\author{\uppercase{Marek Dwornik}$^*$, \uppercase{Zolt\'{a}n Keresztes} and \uppercase{L\'{a}szl\'{o} \'{A}. Gergely}}

\address{Departments of Theoretical and Experimental Physics, University of Szeged,\\
Szeged, 6720 D\'{o}m t\'{e}r 9., Hungary\\
$^*$E-mail: marek@titan.physx.u-szeged.hu}

\begin{abstract}
Rotation curves of spiral galaxies are fundamental tools in the study of dark matter. 
Here we test the Bose-Einstein condensate (BEC) dark matter model against rotation curve data of High and Low Surface 
Brightness (HSB and LSB) galaxies, respectively. When the rotational velocities increase over the whole observed range, 
the fit of the BEC model is similar to the one of the Navarro-Frenk-White (NFW) dark matter model.
When however the rotation curves exhibit long flat regions, the NFW profiles provide a slightly better fit.

\end{abstract}


\bodymatter

\section{Introduction}
The observation of galactic rotational curves show that the amount of luminous matter does not suffice to keep the 
curves flat\cite{persic96}. Numerical N-body simulations in the framework of the $\Lambda$CDM model predict a density 
profile (Navarro-Frenk-White) with a central cusp\cite{nfw}. However on the observational side, rotation curves show a 
shallower density distribution.

It has been proposed that galactic dark matter halos could be Bose-Einstein Condensates (BEC) in Ref.~\refcite{sin94, harko07}. In the BEC model, light bosons are in the same quantum ground state, yielding a repulsive interaction among them. 
This prevents the formation of a cuspy central density and predicts a nearly constant density core, in better 
agreement with the observations.

Here we confront the galactic rotation curves of HSB and LSB galaxies with both the BEC and the NFW model.

\section{Dark matter models}

\subsection {Bose-Einstein Condensate}
The mass density distribution of dark matter BEC 
is\cite{harko07}
\begin{equation}
\rho _{DM}\left( r\right) =\rho _{DM}^{(c)}\frac{\sin kr}{kr} ~,  
\label{becdensity}
\end{equation}
where $\rho _{DM}^{(c)}$ is the central
density of the condensate; $\rho _{DM}^{(c)}=\rho _{DM}(0)$. 
The size $R_{DM} $ of the dark matter BEC halo is defined as $\rho (R_{DM})=0$, 
giving $k=\pi/ R_{DM}$.

The mass profile of the galactic halo, defined as $M_{BEC}(r)=4\pi
\int_{0}^{r}\rho _{DM}(r)r^{2}dr$, is $M_{BEC}\left( r\right) =\frac{4\pi
\rho _{DM}^{(c)}}{k^{2}}r\left( \frac{\sin kr}{kr}-\cos kr\right)$. The
velocity profile is obtained as 
\begin{equation}  \label{becvel}
v_{DM}^{2}(r) =\frac{4\pi G\rho _{DM}^{(c)}}{k^{2}} \left( \frac{\sin kr}{kr}
-\cos kr\right) ,
\end{equation} 
where $G$ is the gravitational constant.

\subsection{The Navarro-Frenk-White dark matter profile}
N-body simulations performed in the framework of the $\Lambda$CDM 
model give the mass density profile\cite{nfw}
\begin{equation}
\rho (r)=\frac{\rho _{s}}{\left( r/r_{s}\right) \left( 1+r/r_{s}\right) ^{2}} ~,
\label{nfwdensity}
\end{equation}
where $\rho _{s}$ and $r_{s}$ are the characteristic density and scale radius, respectively.

The mass within a sphere with radius $r=yr_{s}$ is then given by 
\begin{equation}
M_{NFW}(r)=4\pi \rho _{s}r_{s}^{3}\left[ \ln (1+y)-\frac{y}{1+y}\right] 
\label{nfwmass}
\end{equation}
where $y$ is a dimensionless radial coordinate.
The rotational velocity arises as 
\begin{equation} \label{nfwvel}
v_{NFW}^{2}(r)=\frac{GM_{NFW}(r)}{r} .
\end{equation}

\section{The baryonic model}
In the case of LSB galaxies, the baryonic component consists of a thin exponential disk with the circular 
velocity profile\cite{freeman70}
\begin{equation}  \label{diskvel}
v_{d}^{2}(x)=\frac{GM_{D}^{HSB}}{2h^{HSB}}x^{2}(I_{0}K_{0}-I_{1}K_{1}) ,
\end{equation}
where $x=r/h^{HSB}$. $I_{n}$ and $K_{n}$ are the modified Bessel
functions calculated at $x/2$. Finally $M_{D}^{HSB}$ and $h^{HSB}$ are the total mass and the length scale of the
disk, respectively.
In a HSB galaxy, beside the disk component there is a spherically symmetric bulge. 
The rotational velocity distribution of the bulge is
\begin{eqnarray} \label{bulgevel}
v_{b}^{2}(r) &=&\sigma \frac{G\mathcal{N}(D)}{rF_{\odot }}2\pi
\int\limits_{0}^{r}I_{b}(r)rdr,
\end{eqnarray}
where $F_{\odot }\left( D\right) $ is the apparent flux density of the Sun
at a distance $D$, $\sigma$ is the mass-to-light ratio of the bulge and $I_{b}(r)$ is the 
S\'{e}rsic function\cite{sersic68}. 
Finally $\mathcal{N}(D)=4.4684\times 10^{-35}\left(D/1\;\mathrm{Mpc}\right)^{-2} \;\mathrm{m}^{-2}\;\mathrm{arcsec}
^{2}.$

\section{Confronting the model with rotation curve data}
We tested the validity of the BEC model by fitting the rotation curves of 2 HSB galaxies\cite{palunas} and 
2 LSB galaxies\cite{deblok} with the sum of the predicted velocity profiles (\ref{becvel}), 
(\ref{diskvel}) and (\ref{bulgevel}). The results are compared with the sum of (\ref{diskvel}), (\ref{bulgevel}) 
and (\ref{nfwvel}). 

\begin{figure}[!ht]%
\begin{center}
  \parbox{2.1in}{\epsfig{figure=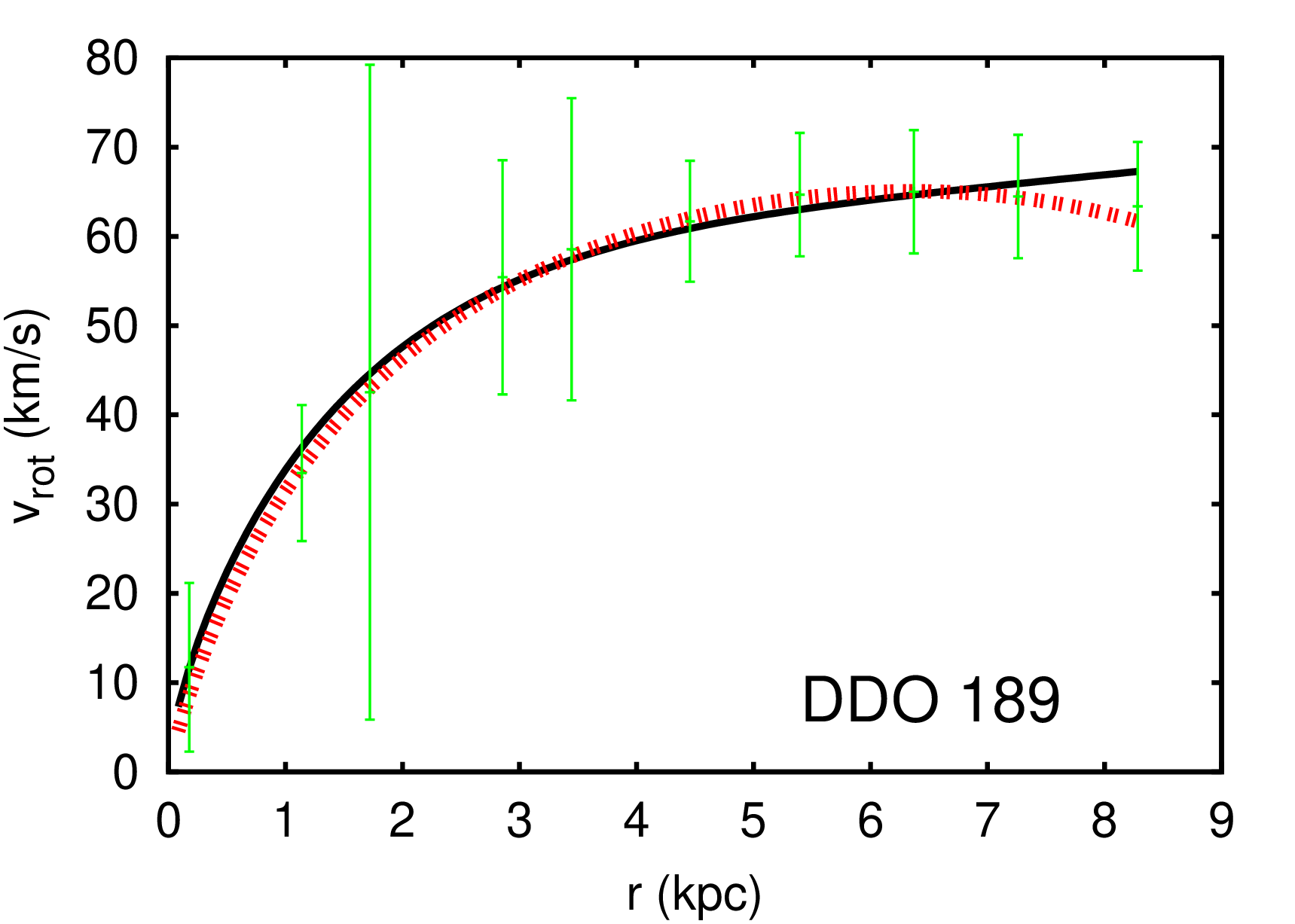,width=1.5in, angle=270}}
  \hspace*{5pt}
  \parbox{2.1in}{\epsfig{figure=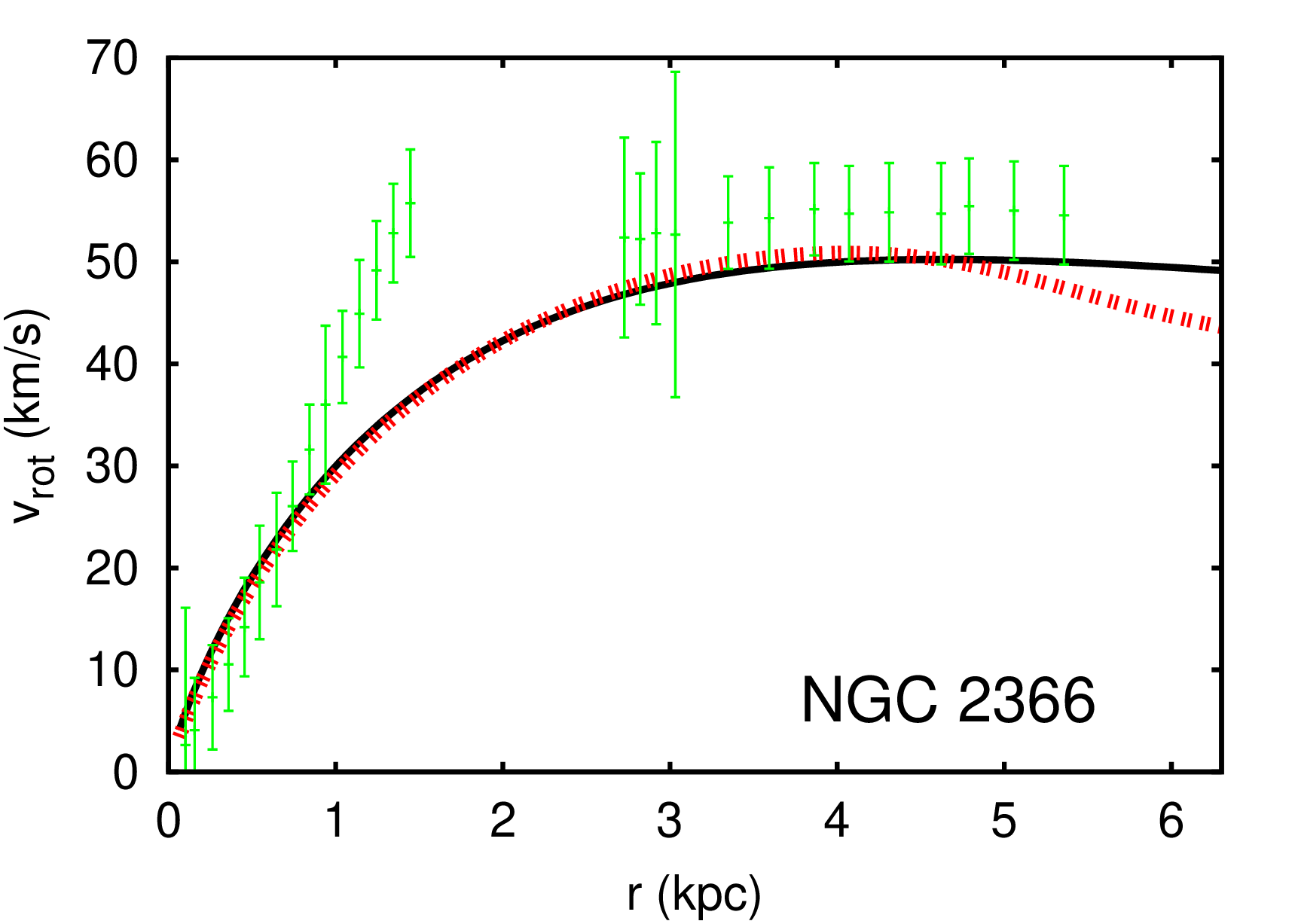,width=1.5in, angle=270}}
  \caption{Best fit curves of the two LSB galaxies: dotted red lines for the BEC+baryonic, 
  solid black lines for the NFW+baryonic model. For the DDO 189 galaxy both models give comparable fits, 
  while in the case of NGC 2366 galaxy, the NFW model gives a slightly better fit.}
  \label{fig1}
\end{center}
\end{figure}

\begin{figure}[!ht]%
\begin{center}
  \parbox{2.1in}{\epsfig{figure=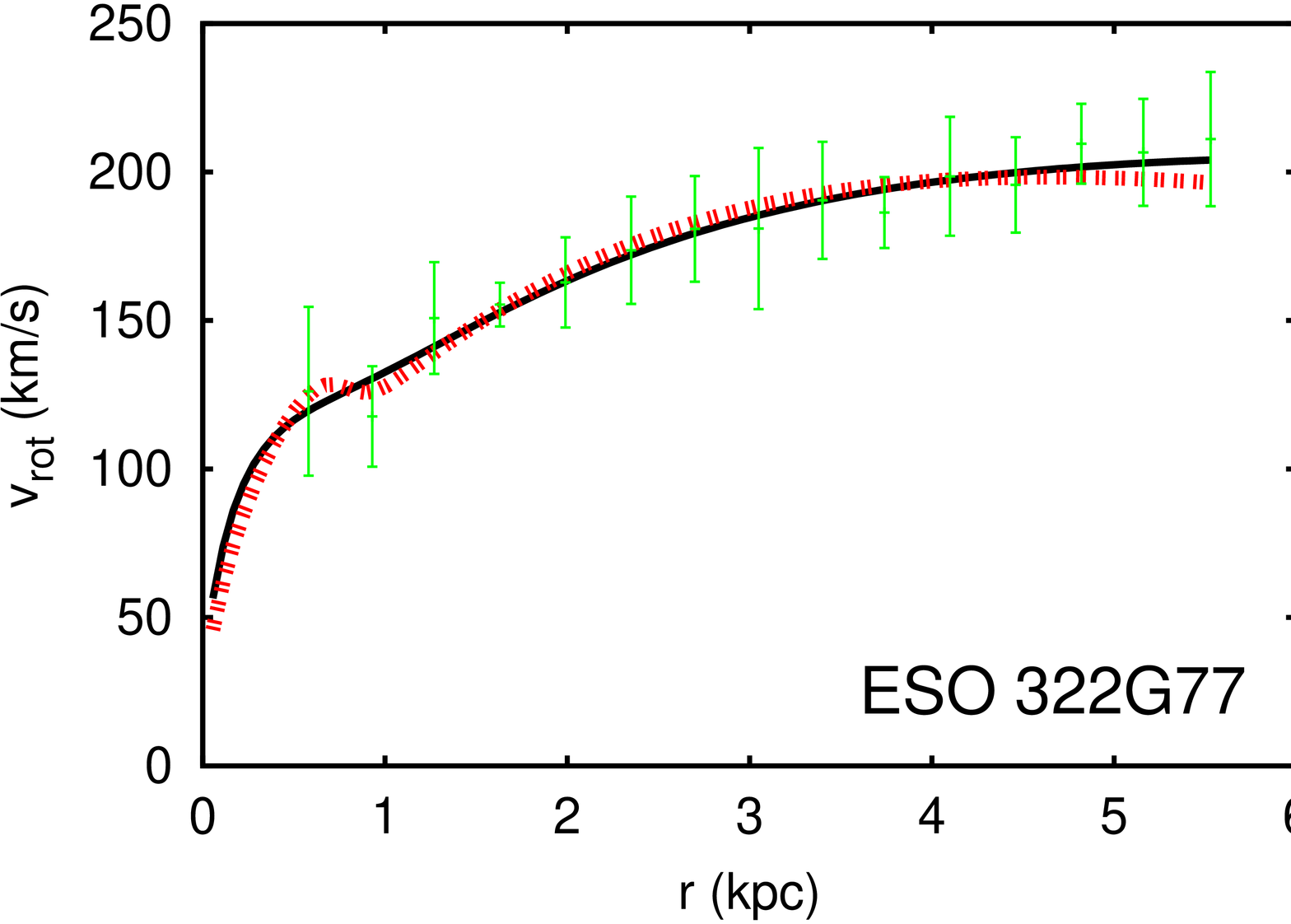,width=1.5in, angle=270}}
  \hspace*{5pt}
  \parbox{2.1in}{\epsfig{figure=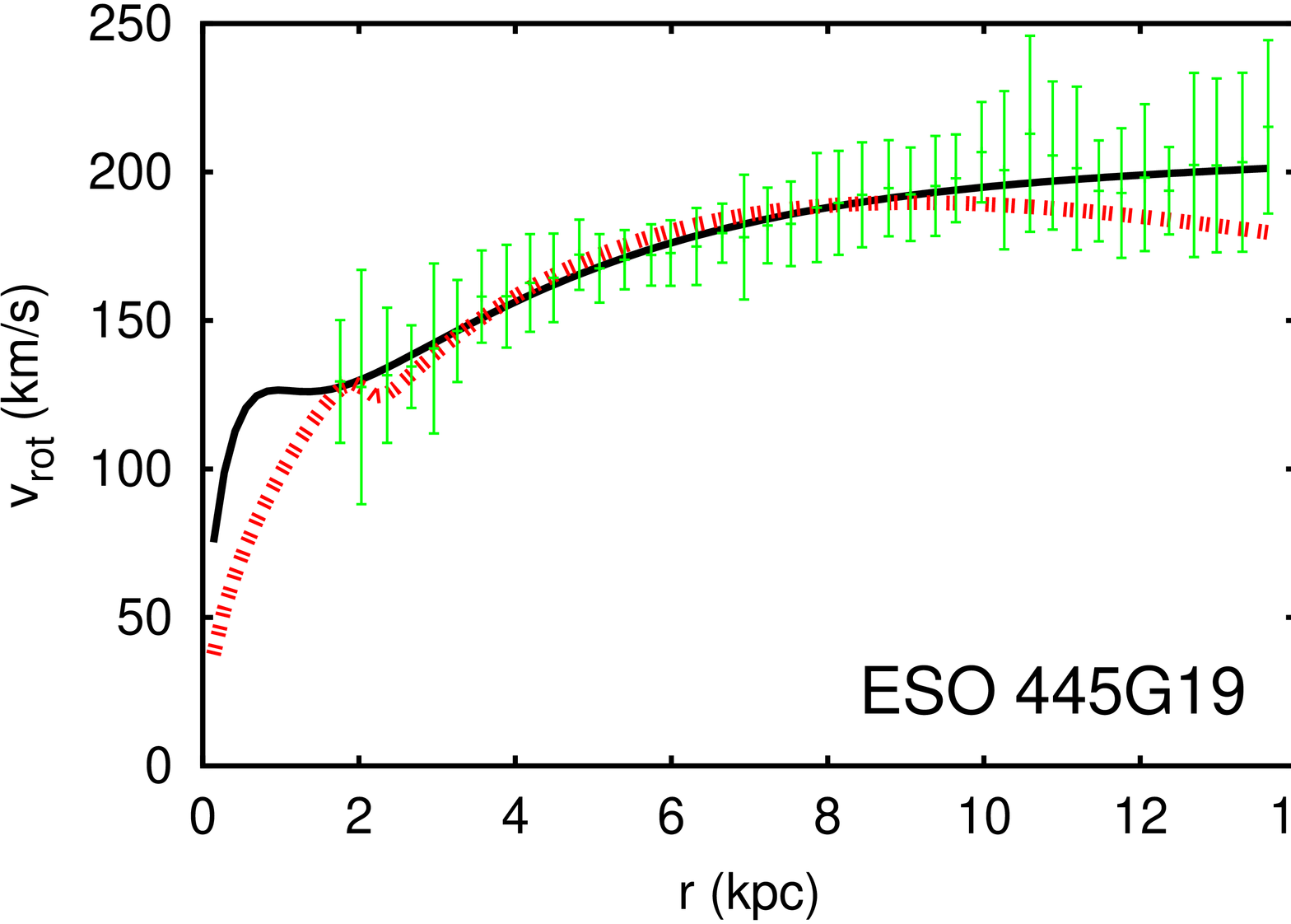,width=1.5in, angle=270}}
  \caption{Best fit curves of the two HSB galaxies: dotted red lines for the BEC+baryonic, 
  solid black lines for the NFW+baryonic model. For the ESO 445G19 galaxy both models give comparable fits, 
  while for the ESO 445G19 galaxy, the NFW model gives a slightly better fit.}
  \label{fig2}
\end{center}
\end{figure}
We determined the best-fit parameters with a $\chi ^{2}$ minimization technique for both the BEC+baryonic and 
NFW+baryonic models. The fitted curves are shown on Figs.~\ref{fig1} and~\ref{fig2}.
For the ESO 322G77 (HSB) and DDO 189 (LSB) galaxies, where the rotational velocities increase over the whole observed 
region, the quality of the fits were comparable for the two models. Similar results were obtained in 
Ref.~\refcite{prague} for the NGC 3274 (LSB) galaxy. However in the case of the 
ESO445G19 (HSB) and NGC2366 galaxies (LSB), the fits of the NFW model were slightly better.

\bibliographystyle{ws-procs975x65}
\bibliography{mdwornik_mg13}

\end{document}